\documentclass[aps,prl,twocolumn,superscriptaddress,nofootinbib]{revtex4-1}

\usepackage{latexsym}
\usepackage{amsmath}
\usepackage{amssymb}
\usepackage{amsfonts}
\usepackage{bm}
\RequirePackage{lineno}

\usepackage{color}
\definecolor{purple}{rgb}{0.5,0,0.5}
\definecolor{blue}{rgb}{0.0,0,0.9}
\definecolor{prdblue}{rgb}{0.133,0.118,0.498}
\usepackage[colorlinks=true, pdfstartview=FitV, linkcolor=prdblue, citecolor= prdblue, urlcolor=prdblue]{hyperref}

\usepackage{supertabular} 
\usepackage{placeins}
\usepackage{epsfig}
\usepackage{graphicx}
\usepackage{hyperref}

\hypersetup{
  breaklinks=true,
  colorlinks = true,
  linkcolor = blue,
  anchorcolor = blue,
  citecolor = blue,
  filecolor = blue,
  pagecolor = blue,
  urlcolor = blue
}
\hypersetup{
  bookmarks=true,
  bookmarksnumbered=true,
  bookmarkstype=toc,
  linktocpage=true
}

\begin{document}

\modulolinenumbers[2]
%\linenumbers

\setlength{\oddsidemargin}{-0.5cm} \addtolength{\topmargin}{15mm}

\title{\boldmath Study of the Semileptonic Decay $\Lambda_c^+\rightarrow \Lambda e^+\nu_e$}
\author{
  \small
M.~Ablikim$^{1}$, M.~N.~Achasov$^{11,b}$, P.~Adlarson$^{70}$, M.~Albrecht$^{4}$, R.~Aliberti$^{31}$, A.~Amoroso$^{69A,69C}$, M.~R.~An$^{35}$, Q.~An$^{66,53}$, X.~H.~Bai$^{61}$, Y.~Bai$^{52}$
% ...
   \\
   \vspace{0.2cm}
   (BESIII Collaboration)\\
   \vspace{0.2cm} {\it
$^{1}$ Institute of High Energy Physics, Beijing 100049, People's Republic of China\\
% ...
$^{4}$ Bochum Ruhr-University, D-44780 Bochum, Germany\\
% ...
$^{11}$ G.I. Budker Institute of Nuclear Physics SB RAS (BINP), Novosibirsk 630090, Russia\\
% ...
$^{31}$ Johannes Gutenberg University of Mainz, Johann-Joachim-Becher-Weg 45, D-55099 Mainz, Germany\\
% ...
$^{35}$ Liaoning Normal University, Dalian 116029, People's Republic of China\\
% ...
$^{52}$ Southeast University, Nanjing 211100, People's Republic of China\\
$^{53}$ State Key Laboratory of Particle Detection and Electronics, Beijing 100049, Hefei 230026, People's Republic of China\\
% ...
$^{61}$ University of Jinan, Jinan 250022, People's Republic of China\\
% ...
$^{66}$ University of Science and Technology of China, Hefei 230026, People's Republic of China\\
% ...
$^{69}$ University of Turin and INFN, (A)University of Turin, I-10125, Turin, Italy; (B)University of Eastern Piedmont, I-15121, Alessandria, Italy; (C)INFN, I-10125, Turin, Italy\\
$^{70}$ Uppsala University, Box 516, SE-75120 Uppsala, Sweden\\
% ...
$^{b}$ Also at the Novosibirsk State University, Novosibirsk, 630090, Russia\\
% ...
%$^{m}$ \textcolor{red}{Also at Renmin University of China, Beijing 100872, People's Republic of China}\\
   \vspace{0.4cm}
}
}

%%%%%%%%%%%%%%%%%%%%%%%%%%%%%%%%%%%%%%%%%%%%%%%%%%%%%%%%%%%%%%%%%%%%%%%%%%%%%%%%%%%%%%%%%%
\begin{abstract}
The study of the Cabibbo-favored semileptonic decay
$\Lambda_c^+\rightarrow \Lambda e^+\nu_e$ is reported using a $4.5~\mathrm{fb}^{-1}$ data sample of $e^+e^-$
annihilations collected at center-of-mass energies ranging from 4.600~GeV to
4.699~GeV with the BESIII detector at the BEPCII collider. The branching fraction of the decay is measured to be 
$\mathcal{B}(\Lambda_c^+\rightarrow \Lambda e^+\nu_e)=(3.56\pm0.11_{\rm stat.}\pm0.07_{\rm syst.})\%$, which is the most precise measurement to date. Furthermore, we perform an investigation of the internal dynamics in $\Lambda_c^+\rightarrow \Lambda e^+\nu_e$. We provide the first direct comparisons of the differential decay rate and form factors with those predicted from lattice quantum chromodynamics (LQCD) calculations.  Combining the measured branching fraction with a $q^2$-integrated rate predicted by LQCD, we determine $|V_{cs}|=0.936\pm0.017_{\mathcal{B}}\pm0.024_{\rm LQCD}\pm0.007_{\tau_{\Lambda_c}}$.
\end{abstract}
\pacs{13.30.Ce, 14.65.Dw}% PACS, the Physics and Astronomy Classification Scheme.

\maketitle

%%%%%%%%%%%%%%%%%%%%%%%%%%%%%%%%%%%%%%%%%%%%%%%%%%%%%%%%%%%%%%%%
%%%%%     Introduction       Part                  %%%%%%%%%%%%%
%%%%%%%%%%%%%%%%%%%%%%%%%%%%%%%%%%%%%%%%%%%%%%%%%%%%%%%%%%%%%%%%
The study of $\Lambda_c^+$ semileptonic\,(SL) decays provides valuable
information about weak and strong interactions in baryons containing a 
heavy quark~\footnote{Throughout this Letter, charge-conjugate modes are implied unless explicitly noted.}.
Measurement of the SL decay
rate of the $\Lambda^+_c$ can help elucidate the role of nonperturbative effects in
strong interactions~\cite{Revm67_893}. In particular, the Cabibbo-favored (CF) SL decay $\Lambda_c^+\rightarrow \Lambda \ell^+\nu_{\ell}$ ($\ell=e, \mu$), which is the dominant component in 
$\Lambda_c^+$ SL decays~\cite{pdg2020}, is the most interesting to measure.
Its decay rate depends on the weak quark mixing Cabibbo-Kobayashi-Maskawa (CKM) matrix~\cite{CKM} element $|V_{cs}|$ and strong interaction effects parameterized by form factors describing the hadronic transition between the initial and the final baryons. 
Measurement of $|V_{cs}|$ via $\Lambda_c^+\rightarrow \Lambda \ell^+\nu_{\ell}$ is an important consistency test for the standard model (SM) and a probe for new physics beyond the SM~\cite{Revm67_893,EPJC80_113}, complementary to $D$ meson analyses.

In recent years, significant progress has been achieved in the study of $\Lambda_c^+$ SL decays, both experimentally~\cite{PRL115_221805,PLB767_42,PRL121_251801} and theoretically~\cite{PRC72_035201,CPC42_093101,PRD93_034008,EPJC76_628,PRD95_053005,PRD80_074011,PRL118_082001,PLB792_214,PRD101_094017,PRD104_013005,PRD80_096007,PRD90_114033,PRD93_014021,PRD93_056008,PRD97_034511,2107.13084,2107.13140,TurJ}.
In 2015, the BESIII collaboration reported the first measurement of the absolute branching fraction (BF) for $\Lambda^+_c\rightarrow \Lambda e^+\nu_{e}$ with $\mathcal B({\Lambda^+_c\rightarrow \Lambda e^+\nu_e})=(3.63\pm0.38_{\rm stat.}\pm0.20_{\rm syst.})\%$~\cite{PRL115_221805}. The absolute BF for $\Lambda^+_c\rightarrow \Lambda \mu^+\nu_{\mu}$~\cite{PLB767_42} was reported later. 
These BESIII measurements motivated the  
first LQCD calculation on the CF SL decay $\Lambda_c^+\rightarrow \Lambda \ell^+\nu_{\ell}$ in 2017, with a predicted BF $\mathcal{B}(\Lambda_c^+\rightarrow \Lambda e^+\nu_e)=(3.80\pm0.19_{\rm LQCD}\pm0.11_{\tau_{\Lambda_c^+}})\%$~\cite{PRL118_082001} which is consistent with the BESIII result.
Reference~\cite{PRL118_082001} also investigated the internal dynamics and predicted the form factors and differential decay rates in $\Lambda_c^+\rightarrow \Lambda e^+\nu_{e}$ decay. This was followed by a series of other LQCD calculations~\cite{PRD97_034511,2107.13084,2107.13140,TurJ} in this sector. This theoretical work is important for understanding the decay mechanism in SL decays in the charmed baryon sector.
However, there is no direct experimental data for testing and calibrating calculations of differential decay rates and form factors. 
Experimental studies of the dynamics in $\Lambda_c^+\rightarrow \Lambda e^+\nu_e$ can provide these information.

In this Letter, we report an improved measurement of the absolute BF of $\Lambda_c^+\rightarrow \Lambda e^+\nu_e$ using data sets collected at the BESIII at the center-of-mass energies  $\sqrt{s}=4.600$, 4.612, 4.628, 4.641, 4.661, 4.682, and 4.699 GeV.  The total integrated luminosity for these data samples is $4.5~\mathrm{fb}^{-1}$~\cite{lum_4600,lum_new}, which includes and is about 7 times larger than that used in Ref.~\cite{PRL115_221805}.
Furthermore, for the first time, we measure the differential decay rate and form factors in $\Lambda_c^+\rightarrow \Lambda e^+\nu_e$ and provide the direct comparisons with those 
predicted by the LQCD calculations~\cite{PRL118_082001}. %which is the urgent and critical for testing the LQCD calculations.

%%%%%%%%%%%%%%%%%%%%%%%%%%%%%%%%%%%%%%%%%%%%%%%%%%%%%%%%%%%%%%%%
%%%%%     Detector and software Part               %%%%%%%%%%%%%
%%%%%%%%%%%%%%%%%%%%%%%%%%%%%%%%%%%%%%%%%%%%%%%%%%%%%%%%%%%%%%%%
Details about the BESIII detector design and performance are provided in Ref.~\cite{Ablikim:2009aa}.
A {\sc geant4}-based~\cite{geant4} Monte Carlo (MC) detector simulation is used to determine signal detection efficiencies and
to estimate potential backgrounds. Signal MC samples of $e^+e^-\rightarrow \Lambda_c^+\bar{\Lambda}_c^-$ with one
$\Lambda_c^+$ baryon decaying to $\Lambda e^+\nu_e$ together with a
$\bar{\Lambda}_c^-$ decaying to the hadronic decay mode used for this analysis are generated by {\sc kkmc}~\cite{kkmc} 
with {\sc evtgen}~\cite{nima462_152}. This includes initial-state radiation~\cite{SJNP41_466} and final-state radiation~\cite{plb303_163} effects.
The signal MC sample of the SL decay $\Lambda_c^+\rightarrow \Lambda e^+\nu_e$ is generated using the form factors measured in this work.
To study the backgrounds, an inclusive MC sample consisting of open-charm states, radiative
return to charmonium(like) $\psi$ states at lower masses, and continuum processes of $q\bar{q}$ ($q=u$, $d$, and $s$), along with Bhabha scattering, $\mu^+\mu^-$, $\tau^+\tau^-$, 
and $\gamma\gamma$ events are generated.
All known decay modes of open-charm and $\psi$ states are
simulated using BFs specified by the Particle Data Group (PDG)~\cite{pdg2020}, while the
remaining unknown $\psi$ decays are modeled with {\sc lundcharm}~\cite{lundcharm,lundcharm2}.

%%%%%%%%%%%%%%%%%%%%%%%%%%%%%%%%%%%%%%%%%%%%%%%%%%%%%%%%%%%%%%%%
%%%%%            Physics Analysis                  %%%%%%%%%%%%%
%%%%%%%%%%%%%%%%%%%%%%%%%%%%%%%%%%%%%%%%%%%%%%%%%%%%%%%%%%%%%%%%
We select the ``single-tag'' (ST) and ``double-tag'' (DT) samples as described in Refs.~\cite{PRL115_221805,PLB767_42}.  STs are $\bar{\Lambda}^-_c$ baryons reconstructed from their daughter particles in one of fourteen hadronic decays as used in Ref.~\cite{pKev}. DTs are events with a ST and a $\Lambda^+_c$ baryon reconstructed as $\Lambda e^+\nu_e$.  
The ST $\bar{\Lambda}^-_c$ signals are identified using the beam
constrained mass: 
 $$M_{\rm BC}=\sqrt{(\sqrt{s}/2)^2-|\overrightarrow{p}_{\bar{\Lambda}^-_c}|^2},$$
where $\overrightarrow{p}_{\bar{\Lambda}^-_c}$ is the measured momentum of the ST $\bar{\Lambda}^-_c$.
A kinematic variable $\Delta E=E_{\rm beam}-E_{\bar{\Lambda}^-_c}$ is required to improve the signal significance for ST $\bar{\Lambda}^-_c$ baryons. 
If an event satisfies more than one $\bar{\Lambda}_c^-$ tags, only the tag with the minimum $|\Delta E|$ is kept to avoid the double counting among STs with the same final state. 
The $\Delta E$ requirements, $M_{\rm BC}$ distributions and their ST yields are documented in Ref.~\cite{pKev}. 
The total ST yield reconstructed with fourteen ST modes is $N_{\rm ST}=122\,268\pm474$, where the uncertainty is calculated by the weighted average according to the fit results of each tag mode.

\begin{figure}[tp!]
\begin{center}
   \begin{minipage}[t]{8cm}
   \includegraphics[width=\linewidth]{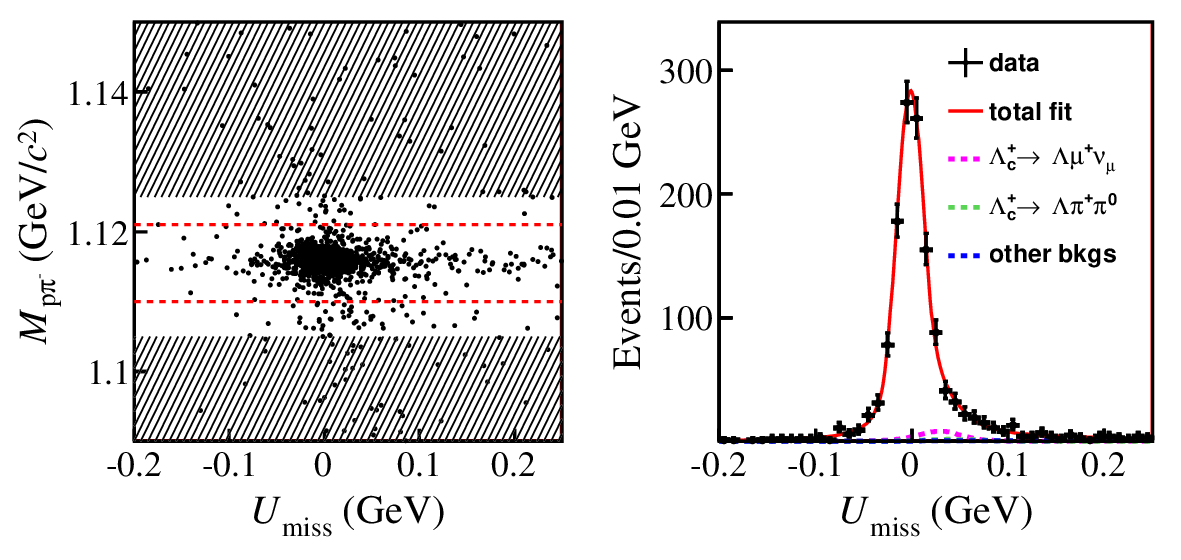}
   \end{minipage}   
   \caption{ (left) The $M_{p\pi^-}$ versus $U_{\rm miss}$ distribution for the $\Lambda_c^+\rightarrow \Lambda e^+\nu_e$
   candidates. The area between the dashed lines denotes the $\Lambda$ signal region and the hatched areas indicate the $\Lambda$ sideband regions. (right) Projected $U_{\rm miss}$ distribution within the $\Lambda$ signal region together with the fit. }
\label{fig:umiss_data_sig}
\end{center}
\end{figure}

Candidates for $\Lambda^+_c\rightarrow \Lambda e^+\nu_e$ are
selected from the remaining tracks recoiling against the ST
$\bar{\Lambda}^-_c$ candidates in the signal mass window. 
To select the $\Lambda$, the same
criteria as those used in the ST selection are applied~\cite{pKev}. Detection and reconstruction of the positron follow the procedures in Ref.~\cite{PRL115_221805}.  
As the neutrino is not detected, we employ the kinematic variable
$U_{\rm miss}=E_{\rm miss}-c|\vec{p}_{\rm miss}|$ to obtain information on the neutrino, where $E_{\rm
miss}$ and $\vec{p}_{\rm miss}$ are the missing energy and momentum
carried by the neutrino, respectively, and are defined in the same way as in Ref.~\cite{PRL115_221805}.
Figure~\ref{fig:umiss_data_sig} (left) shows the distribution of
$M_{p\pi^-}$ versus $U_{\rm miss}$ for the $\Lambda^+_c\to \Lambda
e^+\nu_e$ candidates in data. A cluster of the events is located around
the intersection of the $\Lambda$ mass and $\Lambda e^+\nu_e$ signal
region near $U_{\rm miss} \simeq 0$. Requiring $M_{p\pi^-}$ to be within the $\Lambda$ signal
region, we project the distribution onto the $U_{\rm miss}$ axis, as
shown in Fig.~\ref{fig:umiss_data_sig}(right). 
To obtain the number of signal events, the $U_{\rm miss}$ distribution is fitted with a signal function $f$ which consists of 
a Gaussian to describe the core of the $U_{\rm miss}$
distribution and two power-law tails to account for initial and final state radiations~\cite{PRD79_052010,PRL115_221805}.
Two MC-simulated background shapes are used to describe the peaking backgrounds 
from $\Lambda_c^+\rightarrow \Lambda\pi^+\pi^0$ and $\Lambda_c^+\rightarrow\Lambda\mu^+\nu_{\mu}$, 
and a MC-simulated non-resonant background shape is used to describe the continuous background. 
From the fit, we obtain the yield of $\Lambda_c^+\rightarrow \Lambda e^+\nu_e$ decay $N^{\rm DT}=1253\pm39$, where the uncertainty is statistical only.

%%%%%%%%%%%%%%%%%%%%%%%%%%%%%
%%%%%%%%%%%%%%%%%%%%%%%%%%%%%

The absolute BF for $\Lambda_c^+\rightarrow \Lambda e^+\nu_e$ is
determined by
\begin{equation} \mathcal{B}(\Lambda_c^+\rightarrow \Lambda e^+\nu_e)=\frac{N^{\rm DT}}{N^{\rm ST}\times\varepsilon_{\rm semi}},
\label{eq:branch}
\end{equation}
where $\varepsilon_{\rm semi}=(\sum_{ij} N_{ij}^{\rm ST}\times\epsilon_{ij}^{\rm DT}/\epsilon_{ij}^{\rm ST})/N^{\rm ST}=0.2876$ is the average efficiency for detecting the $\Lambda_c^+\rightarrow \Lambda
e^+\nu_e$ decay in ST events~\cite{PRL115_221805}, and $i$ and $j$ represent the ST modes and
the data samples at different center-of-mass energies, respectively. The parameters $N_{ij}^{\rm ST}$, $\epsilon_{ij}^{\rm ST}$ and $\epsilon_{ij}^{\rm DT}$ are the ST yields, ST and DT efficiencies, respectively. 
Inserting the values of $N^{\rm DT}$, $N^{\rm ST}$, and $\epsilon_{\rm semi}$ in Eq.~(\ref{eq:branch}), we measure $\mathcal B({\Lambda^+_c\rightarrow
\Lambda e^+\nu_e})=(3.56\pm0.11\pm0.07)\%$, where the first uncertainty is statistical, and the second is systematic as described below.

With the DT technique, the BF measurements are insensitive to the systematic uncertainties of the ST selection. The remaining systematic uncertainties in the BF measurement are now described. 
The uncertainties of the $e^+$ tracking and particle identification (PID) efficiencies are determined to be 0.4\% and 0.5\%  using radiative Bhabha events.
The uncertainty due to  $\Lambda$ reconstruction is determined to be 0.2\%, using $J/\psi\rightarrow pK^-\bar{\Lambda}$ and $J/\psi\rightarrow \Lambda\bar{\Lambda}$ control samples.
The uncertainty associated with the simulation of the SL signal model is estimated to be 0.6\% by varying the input form-factor parameters, determined in this work, by one standard deviation. 
The ST yield (1.0\%) is evaluated by using alternative signal shapes in fits to the $M_{\rm BC}$ spectra.  
We also considered the following systematic uncertainties: fit to the
$U_{\rm miss}$ distribution (1.0\%) estimated by using alternative
signal shapes and background shapes, the quoted BF for $\Lambda\rightarrow p\pi^-$
(0.8\%), and the MC statistics (0.8\%).
Adding these contributions in quadrature gives a total systematic uncertainty of 2.0\% for the $\mathcal{B}(\Lambda_c^+\rightarrow \Lambda e^+\nu_e)$ measurement.

The differential decay rate of $\Lambda^+_c\rightarrow \Lambda e^+\nu_e$ can be expressed using four variables:
$e^+\nu_e$ mass-squared ($q^2$), the angle between the proton momentum  in the $\Lambda$ rest frame and the $\Lambda$ momentum in the $\Lambda_c$ rest frame ($\theta_p$),
the angle between the positron momentum in the decay of $W^+\rightarrow e^+\nu_e$ rest frame and the $\Lambda$ momentum in the $\Lambda_c$ rest frame ($\theta_e$), and the acoplanarity angle between the $\Lambda$ and $W^+$ decay planes ($\chi$). It can be described in terms of helicity amplitudes $H_{\lambda_{\Lambda}\lambda_{W}}$ by~\cite{PLB275_495,PRL94_191801,EPJC59_27}:
\small
\begin{eqnarray}
&& \frac{d^4\Gamma}{dq^2d{\rm cos}\theta_{e}d{\rm cos}\theta_{p}d\chi} = \frac{G^2_F|V_{cs}|^2}{2(2\pi)^4}\cdot\frac{Pq^2}{24M^2_{\Lambda_c}} \times  ~~~~~~~ \nonumber \\
&&~~~~\left\{\frac{3}{8}(1-{\rm cos}\theta_{e})^2|H_{\frac{1}{2}1}|^2(1+\alpha_{\Lambda} {\rm cos}\theta_{p})\right. \nonumber \\
&&~~~+\frac{3}{8}(1+{\rm cos}\theta_{e})^2|H_{-\frac{1}{2}-1}|^2(1-\alpha_{\Lambda} {\rm cos}\theta_{p})  \nonumber \\
&&~~~+\frac{3}{4}{\rm sin}^2\theta_{e}[|H_{\frac{1}{2}0}|^2(1+\alpha_{\Lambda}{\rm cos}\theta_{p})+|H_{-\frac{1}{2}0}|^2(1-\alpha_{\Lambda}{\rm cos}\theta_{p})] \nonumber \\
&&~~~+\frac{3}{2\sqrt{2}}\alpha_{\Lambda}{\rm cos}\chi {\rm sin}\theta_{e} {\rm sin}\theta_{p}\times \nonumber \\
&&~~~~~~~~~\left. [(1-{\rm cos}\theta_{e})H_{-\frac{1}{2}0}H_{\frac{1}{2}1}+(1+{\rm cos}\theta_{e})H_{\frac{1}{2}0}H_{-\frac{1}{2}-1}] \right\},  
\label{eq:decayrate}
\end{eqnarray}
\normalsize
where $H_{\lambda_{\Lambda}\lambda_{W}}=H^V_{\lambda_{\Lambda}\lambda_{W}}-H^A_{\lambda_{\Lambda}\lambda_{W}}$ and $H^{V(A)}_{-\lambda_{\Lambda}-\lambda_{W}}=+(-)H^{V(A)}_{\lambda_{\Lambda}\lambda_{W}}$, $G_F$ is the Fermi coupling constant, $|V_{cs}|=0.97320\pm0.00011$~\cite{pdg2020} is a CKM matrix element, $P$ is the magnitude of the $\Lambda$ momentum in
the $\Lambda_c$ rest frame, $M_{\Lambda_c}$ is the mass of $\Lambda_c$~\cite{pdg2020}, and $\alpha_{\Lambda}$ is the $\Lambda\rightarrow p\pi^-$ decay asymmetry parameter~\cite{pdg2020}.

The helicity amplitudes parameterized in Eq.~(\ref{eq:decayrate}) are related to four form factors by~\cite{PRL118_082001}:
\begin{eqnarray}
H^{V}_{\frac{1}{2}1}&=&\sqrt{2Q_-} \, f_{\bot}(q^2), ~~H^{A}_{\frac{1}{2}1}=\sqrt{2Q_+} \, g_{\bot}(q^2), \nonumber \\
H^{V}_{\frac{1}{2}0}&=&\sqrt{Q_-/q^2} \, f_{+}(q^2) \, (M_{{\it \Lambda}_c}+M_{{\it \Lambda}}), \nonumber \\
H^{A}_{\frac{1}{2}0}&=&\sqrt{Q_+/q^2} \, g_{+}(q^2) \, (M_{{\it \Lambda}_c}-M_{{\it \Lambda}}),
\label{eq:Helicity}
\end{eqnarray}
where $Q_{\pm}=(M_{\Lambda_c}\pm M_{\Lambda})^2-q^2$ and $M_{\Lambda}$ is the $\Lambda$ mass~\cite{pdg2020}. The form factors $f_{\bot,+}(q^2)$ and $g_{\bot,+}(q^2)$ are defined following a 
$z$-expansion of the parameters as in Ref.~\cite{PRL118_082001}:  
\begin{eqnarray}
f(q^2)&=&\frac{a_0^{f}}{1-q^2/\left(m_{\rm pole}^{f}\right)^2} \left[1+\alpha^{f}_1\times z(q^2)\right],
\label{eq:formfactor}
\end{eqnarray}
where $m^f_{\rm pole}$ is the pole mass; $a_0^f$, $\alpha_1^f$ are free parameters; 
$z(q^2)=\frac{(\sqrt{t_+-q^2}-\sqrt{t_+-t_0})}{(\sqrt{t_+-q^2}+\sqrt{t_+-t_0})}$ with $t_0=q^2_{\rm max}=(m_{\Lambda_c}-m_{\Lambda})^2$, $t_+=(m_D+m_K)^2$, $m_D=1.870$~GeV/$c^2$ and $m_K=0.494$~GeV/$c^2$. 
The pole masses are $m_{\rm pole}^{f_+,f_{\perp}}=2.112$~GeV/$c^2$ and $m_{\rm pole}^{g_+,g_{\perp}}=2.460$~GeV/$c^2$~\cite{PRL118_082001}.

 For these four form factors, the amplitudes at $q^2=0$ are denoted as $a_0^{f_{\perp}}$, $a_0^{f_{+}}$, $a_0^{g_{\perp}}$, and $a_0^{g_{+}}$, respectively.
As the differential decay rate defined in Eq.~(\ref{eq:decayrate}) has to be normalized,
we can only determine the ratios of these amplitudes and $\alpha^{f}_1$ in the maximum likelihood (ML) fit.
Here, we choose $a_0^{g_{\perp}}$ as the reference since the terms with form factor parameters $a_0^{g_{\perp,+}}$ are dominant compared with that of 
$a_0^{f_{\perp,+}}$ in the $\Lambda_c^+\rightarrow \Lambda e^+\nu_e$ decay. 
Furthermore, we set $\alpha^{g_{\perp}}_1\equiv \alpha^{g_{+}}_1$ and $\alpha^{f_{\perp}}_1\equiv \alpha^{f_{+}}_1$ taking into account that the kinematic dependences of $g_{\perp}(q^2)$ and $g_{+}(q^2)$,  $f_{\perp}(q^2)$ and $f_{+}(q^2)$ as a function of $q^2$ are similar according to the LQCD calculation~\cite{PRL118_082001}. 
Hence, there are five independent free parameters in the ML fit for the differential decay amplitude: $\alpha^{g_{\perp}}_1$, $\alpha^{f_{\perp}}_1$, $r_{f_{+}}=a_0^{f_{+}}/a_0^{g_{\perp}}$, $r_{f_{\perp}}=a_0^{f_{\perp}}/a_0^{g_{\perp}}$, and $r_{g_{+}}=a_0^{g_{+}}/a_0^{g_{\perp}}$.   

\begin{figure}[htbp]
\begin{center}
   \begin{minipage}[t]{8cm}
   \includegraphics[width=\linewidth]{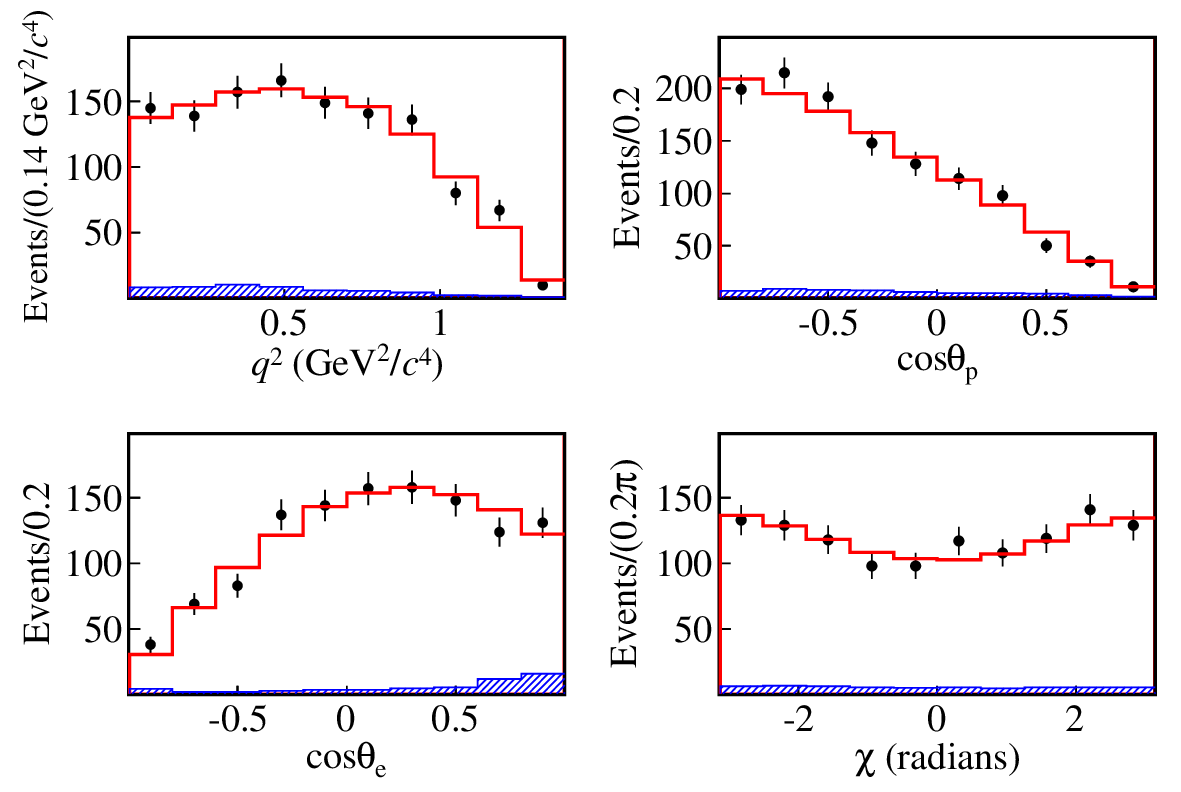}
   \end{minipage}
   \caption{ Projections of the fitted kinematic variables, comparing data (dots with error bars) and MC.  Solid and shadowed histograms are the MC-simulated signal plus background and the MC-simulated background.}
   \label{fig:Form}
\end{center}
\end{figure}

A four-dimensional ML fit is performed as a function of $q^2$, $\cos\theta_e$, $\cos\theta_p$, and $\chi$ for $\Lambda_c^+\rightarrow \Lambda e^+\nu_e$ 
events within $-0.06<U_{\rm miss}<0.06$ GeV.
The projections of the fit onto $q^2$, $\cos\theta_e$, $\cos\theta_p$, and $\chi$ are shown in
Fig.~\ref{fig:Form}. The fit procedure is validated using large inclusive MC samples, 
and the pull distribution of each fitted parameter is consistent with a normal distribution. 
The fitted form-factor parameters of $\alpha_1^{g_{\perp}}$, $\alpha^{f_{\perp}}_1$, $r_{f_+}$, $r_{f_{\perp}}$, $r_{g_{+}}$ are given in Table~\ref{tab:finalresults}. 
The goodness of fit is estimated by using the $\chi^2/{\rm ndof}$ with a similar method as applied in Ref.~\cite{PRD99_011103}, where ${\rm ndof}$ denotes the number of degrees of freedom. The $\chi^2$ is calculated from the comparison between the measured and expected number of events in the four-dimensional space of the kinematic variables $q^2$, $\cos\theta_{e}$, $\cos\theta_p$, and $\chi$ which are initially divided into 2, 3, 3, and 3 bins, respectively. The $\chi^2/{\rm ndof}$ is evaluated to be 0.85.
The systematic uncertainties arise mainly from the uncertainties related to positron
tracking and PID efficiencies, $\Lambda$ reconstruction efficiency, the background normalization and the $\Lambda$ decay parameter. 
The systematic uncertainties arising from the requirements placed on the positron and the $\Lambda$ are estimated by varying the positron tracking and PID efficiencies, and $\Lambda$ reconstruction efficiency by $\pm0.4\%$, $\pm0.5\%$ and $\pm0.2\%$, respectively.
The systematic uncertainty because of the background normalization is estimated by varying its value by $\pm 13\%$ which take into account the uncertainty of the background estimations.
The systematic uncertainty in $\alpha_{\Lambda}$ is evaluated by varying its nominal value by $\pm1\sigma$. 
All of the variations mentioned above will result in differences of the fitted parameters from that under the nominal conditions. 
These differences are assigned as the systematic uncertainties and summarized in Table~\ref{tab:syst}, where the total systematic uncertainty is obtained by adding all contributions in quadrature.

\begin{table*}
\begin{center}
\caption{Measured form-factor parameters, where the first errors are statistical and the second are systematic. The lower table shows the correlation coefficients of statistical and systematic uncertainties between the form factor parameters. } \normalsize
\begin{tabular}
{|l|c|c|c|c|c|c|c|c|} \hline   Parameters & $\alpha^{g_{\perp}}_1$ & $\alpha^{f_{\perp}}_1$ & $r_{f_+}$ & $r_{f_{\perp}}$ & $r_{g_{+}}$ \\
\hline Values &  $1.43\pm2.09\pm0.16$ & $-8.15\pm1.58\pm0.05$ & $1.75\pm0.32\pm0.01$ & $3.62\pm0.65\pm0.02$ & $1.13\pm0.13\pm0.01$ \\
\hline 
\hline Coefficients  & $\alpha^{g_{\perp}}_1$ & $\alpha^{f_{\perp}}_1$ & $r_{f_+}$ & $r_{f_{\perp}}$ & $r_{g_{+}}$ \\
\hline 
 $a_0^{g_{\perp}}$         &  $-0.64$  & ~~~$0.60$ & $-0.66$   & $-0.83$     &  $-0.40$ \\
 $\alpha^{g_{\perp}}_1$ &                &    $-0.63$ & ~~~$0.62$ & ~~~$0.53$ & $-0.33$ \\
 $\alpha^{f_{\perp}}_1$  &                &               & $-0.79$    & $-0.67$     & $-0.07$ \\
 $r_{f_{+}}$                    &                &               &                  & ~~~$0.57$  & $-0.09$ \\
 $r_{f_{\perp}}$              &                &               &                  &                  & ~~~$0.39$ \\
\hline
\end{tabular}
\label{tab:finalresults}
\end{center}
\end{table*}

\begin{table}
\caption{ Systematic uncertainties (in~\%) of the fitted parameters. } 
\begin{center}
%\small
\begin{tabular}
{|l|c|c|c|c|c|c|c|c|} \hline Parameter &  Tracking\&PID\&$\Lambda$ & Normalization & $\alpha_{\Lambda}$ & Total  \\ \hline
$a^{f_{\perp}}_1$  & 0.6  & 0.5  & 0.1 & 0.8 \\ 
$a^{g_{\perp}}_1$  & 6.0  & 7.2  & 2.8 & 9.8 \\ 
$r_{f_+}$          & 0.1  & 0.5  & 0.7 & 0.9 \\ 
$r_{g_{\perp}}$    & 0.3  & 0.1  & 0.6 & 0.7 \\ 
$r_{g_{+}}$        & 0.3  & 1.5  & 0.1 & 1.5 \\ 
\hline
\end{tabular}
\label{tab:syst}
\end{center}
\end{table}

In order to determine the parameter of $a_0^{g_{\perp}}$, the differential decay rate is related to $\mathcal{B}(\Lambda_c^+\rightarrow \Lambda e^+\nu_e)$ measured in this work and the lifetime of $\Lambda_c$ ($\tau_{\Lambda_c}$) by:
\begin{equation}
\int\limits^{q^2_{\rm max}}_{0} \frac{d\Gamma}{dq^2}dq^2=\frac{\mathcal{B}(\Lambda_c^+\rightarrow \Lambda e^+\nu_e)}{\tau_{\Lambda_c}},
\label{eq:a0fperp}
\end{equation}
where $d\Gamma/dq^2$ is the integration of the differential decay rate given in Eq.~(\ref{eq:decayrate}) over the other three kinematic variables and expressed as:
\begin{eqnarray}
\frac{d\Gamma}{dq^2}&=&\frac{G_F^2|V_{cs}|^2}{192\pi^3M^2_{\Lambda_c}}\times Pq^2\times  \nonumber \\
&& \left[ |H_{\frac{1}{2}1}|^2+|H_{-\frac{1}{2}-1}|^2+|H_{\frac{1}{2}0}|^2+|H_{-\frac{1}{2}0}|^2 \right].
\end{eqnarray}
Inserting $\tau_{\Lambda_c}=202.4\pm3.1$~fs~\cite{pdg2020}, the CKM element $|V_{cs}|=0.97320\pm0.00011$~\cite{pdg2020} and the helicity amplitudes parameterized with form factors as in Eq.~(\ref{eq:Helicity}), we determine $a_0^{g_{\perp}}=0.54\pm0.04_{\rm stat.}\pm0.01_{\rm syst.}$. 

\begin{figure}[htbp]
\begin{center}
   \begin{minipage}[t]{8cm}
   \includegraphics[width=\linewidth]{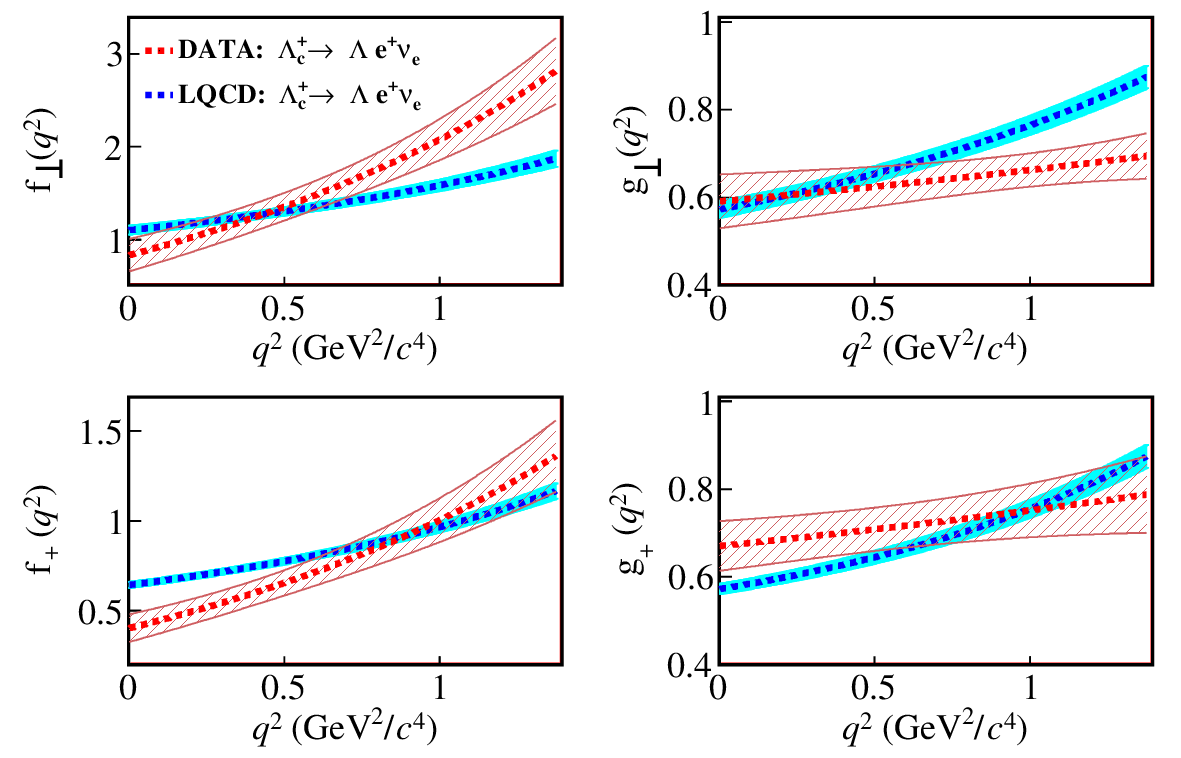}
   \end{minipage}
   \caption{Comparison of form factors with LQCD calculations. The bands show the total uncertainties.} 
   \label{fig:form_Lev}
\end{center}
\end{figure}

\begin{figure}[htbp]
   \begin{center}
   \begin{minipage}[t]{8cm}
   \includegraphics[width=\linewidth]{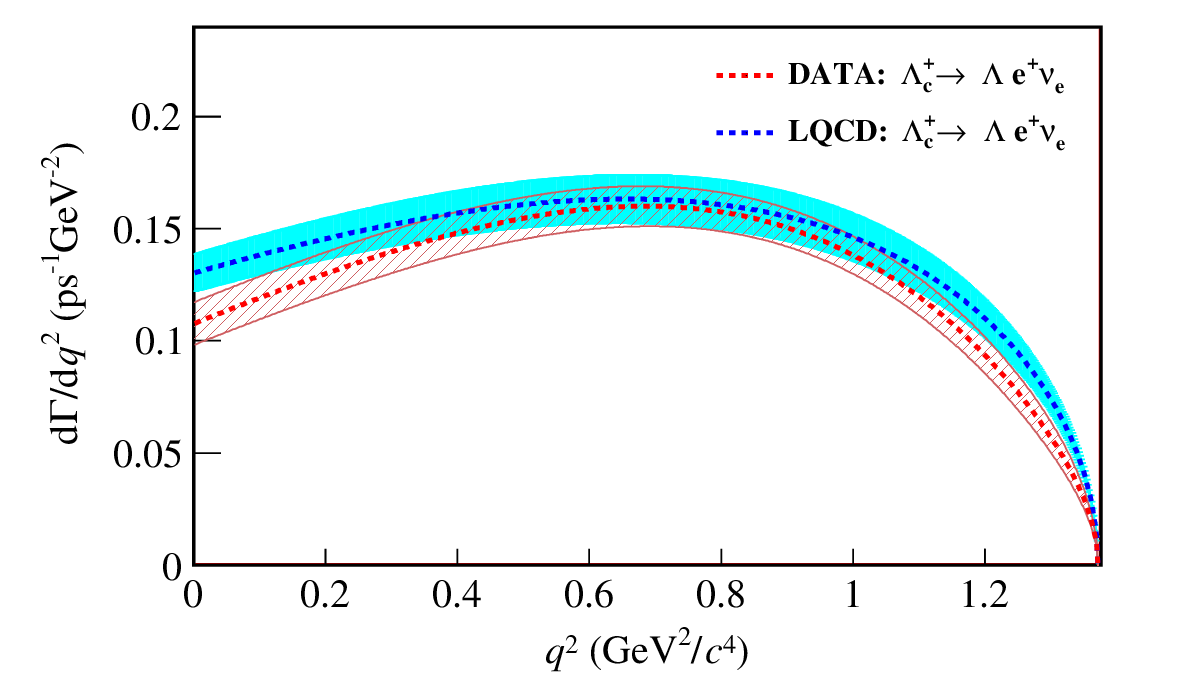}
   \end{minipage}
   \caption{Comparison of the differential decay rates with LQCD predictions. The band show the total uncertainties.}
   \label{fig:Decayrates}
\end{center}
\end{figure}

%With the measured parameters and their correlation coefficients shown in Table~\ref{tab:finalresults}, we obtain the dependences of form factors of $f_{\bot}(q^2)$, $f_+(q^2)$, $g_{\bot}(q^2)$, and $g_{+}(q^2)$, and the differential decay rate as a function of $q^2$ in the SL decay $\Lambda_c^+\rightarrow \Lambda e^+\nu_e$. Figures~\ref{fig:form_Lev} and~\ref{fig:Decayrates} show comparisons of the FFs and the differential decay rate between measurements in this work and that in LQCD calculations. Results of LQCD calculations are obtained using the nominal values listed in Table~IV in Ref.~\cite{PRL118_082001}. 
%\textcolor{red}{We note that the dependences of measured FFs show different kinematic behavior compared to those predicted from LQCD calculations. }
%\textcolor{blue}{The $q^2$ dependences of the measured FFs are different from those predicted from LQCD calculations.}
%However, no clear difference is observed within uncertainties for the resulting differential decay rate. The following reason accounts for this situation. For $f_{+}(q^2)$ and $f_{\perp}(q^2)$ our measurement seems to have steeper slope than those from LQCD calculations, while on the contrary for $g_{+}(q^2)$ and $g_{\perp}(q^2)$. Therefore the shapes of differential decay rates between measurement and LQCD prediction tend to be similar.

With the measured parameters and their correlation coefficients shown in Table~\ref{tab:finalresults}, we obtain the dependences of form factors of $f_{\bot}(q^2)$, $f_+(q^2)$, $g_{\bot}(q^2)$, and $g_{+}(q^2)$, and the differential decay rate as a function of $q^2$ in the SL decay $\Lambda_c^+\rightarrow \Lambda e^+\nu_e$. Figures~\ref{fig:form_Lev} and~\ref{fig:Decayrates} show comparisons of the FFs and the differential decay rate between measurements in this work and that in LQCD calculations. Results of LQCD calculations are obtained using the nominal values listed in Table~IV in Ref.~\cite{PRL118_082001}. 
We note that the dependences of measured FFs show different kinematic behavior compared to those predicted from LQCD calculations.
In particular, discrepancies can be seen at high $q^2$ regions for $f_{\perp}(q^2)$ and $g_{\perp}(q^2)$, as well as at low $q^2$ regions for $f_{+}(q^2)$. For $f_{+}(q^2)$ and $f_{\perp}(q^2)$, our measurement tends to have steeper slope than those from LQCD calculations, while on the contrary for $g_{+}(q^2)$ and $g_{\perp}(q^2)$.
The corresponding comparison on differential decay rates, shown in Fig.~\ref{fig:Decayrates},  gives fair agreement throughout the $q^2$ region.

In summary, we report an improved measurement of the absolute BF for
$\Lambda^+_c\rightarrow \Lambda e^+\nu_e$, 
$\mathcal B({\Lambda^+_c\rightarrow \Lambda
e^+\nu_e})=(3.56\pm0.11\pm0.07)\%$, based on 4.5~fb$^{-1}$ of data
collected at center-of-mass energies 
ranging from 4.600~GeV to 4.699~GeV with BESIII. 
This work supersedes our previous measurement~\cite{PRL115_221805} and improves the precision of
the world average value more than threefold. Comparisons of the BF from this work and theoretical predictions are shown in Table~\ref{tab:theorybf}. 
The predicted BFs in Refs.~\cite{PRC72_035201,PRD93_034008,EPJC76_628,PRD95_053005,CPC42_093101} differ by more than two standard deviations with respect to the mean value of our measured
$\mathcal B({\Lambda^+_c\rightarrow \Lambda e^+\nu_e})$. Thus, our measurement disfavors these predictions at a confidence level of more than $95\%$. Combing $\mathcal B({\Lambda^+_c\rightarrow \Lambda e^+\nu_e})$ measured in this work, $\tau_{\Lambda_c}$ and the $q^2$-integrated rate predicted by LQCD~\cite{PRL118_082001}, we determine $|V_{cs}|=0.936\pm0.017_{\mathcal{B}}\pm0.024_{\rm LQCD}\pm0.007_{\tau_{\Lambda_c}}$, which is in consistent with $|V_{cs}|=0.939\pm0.038$ measured in $D\rightarrow K\ell\nu_{\ell}$ decays~\cite{pdg2020} within one standard deviation.

Furthermore, by analyzing the dynamics of $\Lambda_c^+\rightarrow \Lambda e^+\nu_e$, we measure the $\Lambda_c^+\rightarrow \Lambda$ decay form factors for the first time. Our results provide the first direct comparisons on the differential decay rates and form factors with those obtained from LQCD calculations. Currently, the statistical uncertainty dominates in the total uncertainty of our result and the measurement can be improved with more experimental data available in the future~\cite{CPC44_040001}. The comparisons of differential decay rate and form factors between experimental measurements and LQCD predictions presented in this work provide important inputs in understanding the SL decays of charmed baryons.
Our results will also help calibrate the calculation of SL decays of other charmed baryons~\cite{2103.07064,PLB823_13675} as well as the $\Lambda_b$~\cite{PRD104_013005,PRD80_096007,PRD90_114033,PRD85_014035,PRD88_014512,PRD92_034503,PRD93_074501,2107.13140}. 

\begin{table}
\caption{ Comparison of $\mathcal{B}(\Lambda_c^+\rightarrow \Lambda e^+\nu_e)$ from theoretical calculations and our measurement. } 
\begin{center}
\resizebox{!}{2.3cm}{
\begin{tabular}
{l|c} \hline\hline 
 &  $\mathcal{B}(\Lambda_c^+\rightarrow \Lambda e^+\nu_{e})$ [\%]  \\ \hline
Constituent quark model (HONR)~\cite{PRC72_035201} & 4.25 \\ 
Light-front approach~\cite{CPC42_093101} & 1.63 \\
Covariant quark model~\cite{PRD93_034008}     & 2.78 \\
Relativistic quark model~\cite{EPJC76_628}            & 3.25 \\
Non-relativistic quark model~\cite{PRD95_053005}     & 3.84 \\
Light-cone sum rule~\cite{PRD80_074011}       & $3.0\pm0.3$ \\
Lattice QCD~\cite{PRL118_082001}     & $3.80\pm0.22$ \\
$SU(3)$~\cite{PLB792_214}                   & $3.6\pm0.4$ \\      
Light-front constituent quark model~\cite{PRD101_094017} & $3.36\pm0.87$ \\
MIT bag model~\cite{PRD101_094017} & 3.48 \\
Light-front quark model~\cite{PRD104_013005} & $4.04\pm0.75$ \\
This work                                     & $3.56\pm0.11\pm0.07$ \\
\hline\hline
\end{tabular}
}
\label{tab:theorybf}
\end{center}
\end{table}

%%%%%%%%%%%%%%%%%%%%%%%%%%%%%%%%%%%%%%%%%%%%%%%%%%%%%%%%%%%%%%%%
%%%%%    acknowledgments       Part                %%%%%%%%%%%%%
%%%%%%%%%%%%%%%%%%%%%%%%%%%%%%%%%%%%%%%%%%%%%%%%%%%%%%%%%%%%%%%%
We thank Wei Wang for useful discussions on the differential decay rate of $\Lambda_c^+\rightarrow \Lambda e^+\nu_e$.
The BESIII collaboration thanks the staff of BEPCII and the IHEP computing center for their strong support. This work is supported in part by National Key R\&D Program of China under Contracts Nos.  2020YFA0406400, 2020YFA0406300; National Natural Science Foundation of China (NSFC) under Contracts Nos. 11635010, 11735014, 11835012, 11935015, 11935016, 11935018, 11961141012, 12022510, 12025502, 12035009, 12035013, 12192260, 12192261, 12192262, 12192263, 12192264, 12192265; the Chinese Academy of Sciences (CAS) Large-Scale Scientific Facility Program; Joint Large-Scale Scientific Facility Funds of the NSFC and CAS under Contract No. U1832207; CAS Key Research Program of Frontier Sciences under Contract No. QYZDJ-SSW-SLH040; 100 Talents Program of CAS; INPAC and Shanghai Key Laboratory for Particle Physics and Cosmology; ERC under Contract No. 758462; European Union's Horizon 2020 research and innovation programme under Marie Sklodowska-Curie grant agreement under Contract No. 894790; German Research Foundation DFG under Contracts Nos. 443159800, Collaborative Research Center CRC 1044, GRK 2149; Istituto Nazionale di Fisica Nucleare, Italy; Ministry of Development of Turkey under Contract No. DPT2006K-120470; National Science and Technology fund; STFC (United Kingdom); The Royal Society, UK under Contracts Nos. DH140054, DH160214; The Swedish Research Council; U.S. Department of Energy under Contract No. DE-FG02-05ER41374.
%%%%%%%%%%%%%%%%%%%%%%%%%%%%%%%%%%%%%%%%%%%%%%%%%%%%%%%%%%%%%%%%
%%%%%    bibliographies       Part                %%%%%%%%%%%%%
%%%%%%%%%%%%%%%%%%%%%%%%%%%%%%%%%%%%%%%%%%%%%%%%%%%%%%%%%%%%%%%%

\end{document}